\documentclass[12pt,letterpaper]{article}
\usepackage{epsfig,rotating,setspace,latexsym,amsmath,epsf,amssymb,amsfonts,bm,theorem,cite,caption,subcaption,enumerate,longtable,accents}
\usepackage{algorithm,algorithmic,graphicx,epsf,authblk,epstopdf,url,color,multirow}

\newtheorem{theorem}{Theorem}

\newtheorem{definition}{Definition}
\newtheorem{remark}{Remark}
\newtheorem{lemma}{Lemma}
\newenvironment{Proof}[1]{\medskip\par\noindent{\bf Proof:\,}\,#1}{{\mbox{\,$\blacksquare$}\par}}

\newcommand{\cq}{{\mathcal{Q}}}
\newcommand{\cR}{{\mathcal{R}}}
\newcommand{\cw}{{\mathcal{W}}}

\allowdisplaybreaks

\begin{document}
	
\title{Private Information Retrieval from Non-Replicated Databases\thanks{This work was supported by NSF Grants CNS 15-26608, CCF 17-13977 and ECCS 18-07348.}}
	
\author{Karim Banawan \qquad Sennur Ulukus\\
	\normalsize Department of Electrical and Computer Engineering\\
	\normalsize University of Maryland, College Park, MD 20742 \\
	\normalsize {\it kbanawan@umd.edu} \qquad {\it ulukus@umd.edu}}
	
\maketitle
	
\vspace*{-0.8cm}

\begin{abstract}
We consider the problem of private information retrieval (PIR) of a single message out of $K$ messages from $N$ non-colluding and non-replicated databases. Different from the majority of the existing literature, which considers the case of replicated databases where  all databases store the same content in the form of all $K$ messages, here, we consider the case of \emph{non-replicated databases} under a special non-replication structure where each database stores $M$ out of $K$ messages and each message is stored across $R$ different databases. This generates an \emph{$R$-regular graph} structure for the storage system where the vertices of the graph are the messages and the edges are the databases. We derive a general upper bound for $M=2$ that depends on the graph structure. We then specialize the problem to storage systems described by two special types of graph structures: \emph{cyclic graphs} and \emph{fully-connected graphs}. We prove that the PIR capacity for the case of cyclic graphs is $\frac{2}{K+1}$, and the PIR capacity for the case of fully-connected graphs is $\min\{\frac{2}{K},\frac{1}{2}\}$. To that end, we propose novel achievable schemes for both graph structures that are capacity-achieving. The central insight in both schemes is to introduce \emph{dependency} in the queries submitted to databases that do not contain the desired message, such that the requests can be \emph{compressed}. In both cases, the results show severe degradation in PIR capacity due to non-replication.
\end{abstract}

\section{Introduction}
Private information retrieval (PIR), introduced in \cite{ChorPIR}, is a canonical problem to study the privacy of users as they download content from public databases. In the classical setting, a user is interested in retrieving a single message (file) out of $K$ messages from $N$ replicated and non-colluding databases, in such a way that no database can know the identity of the user's desired file. The PIR problem has become a vibrant research topic within information theory starting with trailblazing papers \cite{RamchandranPIR, unsynchonizedPIR, YamamotoPIR, VardyConf2015, RazanPIR, JafarPIRBlind, PIR_Globecomm}. In \cite{JafarPIR}, Sun and Jafar introduce the PIR capacity, which is the supremum of the ratio of the number of bits of desired information ($L$) that can be retrieved privately to the total downloaded information. They characterize the PIR capacity of the classical PIR problem to be $C_{\text{PIR}}=(1+\frac{1}{N}+\cdots+\frac{1}{N^{K-1}})^{-1}$. The fundamental limits of many interesting variants of the problem have been investigated in \cite{JafarColluding, arbitraryCollusion, RobustPIR_Razane, symmetricPIR, KarimCoded, arbmsgPIR, MultiroundPIR, codedsymmetric, codedcolluded, codedcolludedJafar,MPIRjournal, codedcolludingZhang, MPIRcodedcolludingZhang, CodeColludeByzantinePIR, BPIRjournal,symmetricByzantine, wang2017linear, tandon2017capacity, wei2017fundamental, kadhe2017private, chen2017capacity, wei2017capacity, wei2017fundamental_partial,KimCache, StorageConstrainedPIR_Wei, SI_Gastpar, sun2017_computation, mirmohseni2017private, PrivateSearch, abdul2017private, StorageConstrainedPIR, KarimAsymmetricPIR, PIR_WTC_II, noisyPIR, SecurePIR, securePIRcapacity, securestoragePIR, XSTPIR, Tian_upload, LiConverse, Staircase_PIR, PIR_lifting, PIR_networks, PIR_cache_edge, PIR_decentralized, TamoISIT, Tamo_journal}.

A common assumption in most of these works is that the entire message set is \emph{replicated} across all databases. This is crucial for constructing capacity-achieving schemes, as in many existing schemes the undesired symbols downloaded from one database are exploited as side information in the remaining databases, and replication is the key that enables downloading any bit from any database and using it as side information at any other database. However, the replication assumption may not be practical in next-generation storage systems and networks. From a storage point of view, message replication is impractical as it incurs high storage cost, especially for storage systems with a large number of messages or files with a large size. From a network structure point of view, in next-generation networks where peer-to-peer (P2P) connections will be prevalent, nodes (i.e., databases) may not necessarily possess the same set of messages. These practical scenarios, which challenge the replication assumption, motivate investigating PIR in \emph{non-replicated} storage systems. In this work, we aim at devising achievable schemes that do not rely on message replication, and at the same time, that are more efficient than the trivial scheme of downloading the contents of all databases. We aim at evaluating the loss in the PIR rate due to non-replication and investigating the interplay between the storage structure and the resulting PIR rate.

A few works have considered relaxing the replication assumption: Reference \cite{KarimCoded} investigates the case when the contents of the databases are encoded via an $(N,K_c)$ MDS code instead of assuming data replication. \cite{KarimCoded} derives the PIR capacity for this setting, which reveals a fundamental tradeoff between storage cost and retrieval cost. Reference \cite{StorageConstrainedPIR} studies the PIR problem from storage constrained databases. In this problem, each database is constrained to store $\mu KL$ uncoded bits with $\mu\leq 1$ (as opposed to $KL$ bits needed in replicated databases). \cite{StorageConstrainedPIR} shows that symmetric batch caching, which was originally introduced for centralized coded caching systems in \cite{Caching_Maddah_Ali}, results in the largest possible PIR rate under storage constraints. This problem is extended to the decentralized setting in \cite{PIR_decentralized}, where each database stores $\mu KL$ bits randomly and independently from any other database. \cite{PIR_decentralized} shows that uniform and random bit selection, which was introduced for decentralized coded caching systems in \cite{decentralized_caching}, results in the largest possible PIR rate under storage constraints.

The work that is most closely related to our work here is \cite{TamoISIT}. The databases in \cite{TamoISIT} store different subsets of the message set. Different from the previous works on non-replication such as \cite{StorageConstrainedPIR, PIR_decentralized}, in \cite{TamoISIT} databases store \emph{full messages} and not portions of every message. In particular, \cite{TamoISIT} investigates the case when every message is replicated across two databases only. This storage system, in this case, can be represented by a graph, in which every two databases are connected via an edge corresponding to the common message. \cite{TamoISIT} proposes an achievable PIR scheme that is immune against colluding databases, that do not form a cycle in the graph. The scheme in \cite{TamoISIT} achieves a retrieval rate of $\frac{1}{N}$. The work in \cite{TamoISIT} highlights some interesting insights about the relation between some combinatorial properties of the graph and the immunity against database collusion. In the extended version of \cite{TamoISIT} in \cite{Tamo_journal}, which has appeared concurrently and independently of our work here, an upper bound is proposed to show that their PIR rate is at most a factor of 2 from the optimal value for regular graphs, and the techniques are extended to larger replication factors.

In this paper, we consider PIR of a single message out of $K$ messages from $N$ non-replicated and non-colluding databases. In our formulation, each message appears in $R$ different databases, and every database stores $M$ different messages. Thus, the storage system is parameterized by $(K,R,M,N)$ such that $KR=MN$, where $K$ is the total number of messages in the system, $R$ is the replication factor of each message, $M$ is the storage constraint of each database, and $N$ is the number of databases. We focus on the case $M=2$. For this case, the storage system can be uniquely specified by an \emph{$R$-regular graph}. In our graph formulation, the messages correspond to the vertices and the databases correspond to the edges. This is in contrast to \cite{TamoISIT}, where $R=2$, and the roles of messages and databases are reversed on the graph. Hence, our graph formulation may be considered as the dual graph formulation to \cite{TamoISIT}. Our goal is to characterize the PIR capacity of this system.

First, we derive a general upper bound on the retrieval rate for storage systems described by $R$-regular graphs. Interestingly, the upper bound depends on the structure of the graph and not only on $(K,R,M,N)$. In particular, the upper bound is related to the longest sequence of databases that cover all of the $K$ messages in the storage system. We specialize the problem further to two classes of graphs, namely, cyclic graphs and fully-connected graphs, where we obtain exact results. In \emph{cyclic graphs}, all vertices form a circle connected by edges. Therefore, each vertex (a message) emanates two edges (two databases), which means that each message is common among two adjacent databases which are arranged in a cycle. Thus, in this case $R=2$, and since $M=2$ in this paper, using $KR=MN$ mentioned above, we have, $K=N$. For this type of graphs, we show that $C_{\text{PIR}}=\frac{2}{K+1}$. The achievable scheme starts from the greedy algorithm of Sun and Jafar \cite{JafarPIR} and then compresses the requests to $K-2$ databases by replacing the individual symbols of the scheme in \cite{JafarPIR} by sum of two messages. This compression necessitates exploiting side information even in databases that do not contain the desired messages. In \emph{fully-connected graphs}, each vertex is connected to all of the remaining $K-1$ vertices. Therefore, each vertex (a message) emanates $K-1$ edges ($K-1$ databases), which means that each message resides in $K-1$ databases. Thus, in this case $R=K-1$, and since $M=2$, from $KR=MN$, we have $N=K(K-1)/2$, i.e., $N=\binom{K}{2}$. That is, all $\binom{K}{2}$ combinations of two messages appear in a different database. In this case, we show that $C_{\text{PIR}}=\min\{\frac{2}{K}, \frac{1}{2}\}$. For $K \geq 4$, for this case, we propose a novel achievable scheme, which is based on retrieving a single weighted sum (with respect to sufficiently large field) of two symbols from every database. For the comparable cases with \cite{TamoISIT}, our scheme outperforms their scheme in terms of the PIR rate. We note that, in both cyclic and fully-connected graph cases, the PIR capacity converges to zero as $N \rightarrow \infty$, which implies a severe degradation in the PIR efficiency due to non-replication. 

Finally, we show an example for a storage system with $M=3$. We provide a novel achievable scheme that uses \emph{processed side information} and outperforms the scheme in \cite{TamoISIT}.

\section{Problem Formulation}
Consider the problem of PIR from $N$ non-replicated and non-colluding databases. We denote the databases by $\mathcal{D}=\{D_1, D_2, \cdots, D_N\}$. The storage system stores $K$ messages in total, each message is stored across $R$ different databases, i.e., $R$ is the repetition factor for every message, and each database stores locally $M$ different messages. We denote the message set by $\cw=\{W_1, W_2, \cdots, W_K\}$. Each message $W_k \in \mathbb{F}_q^L$ is a vector of length $L$ picked in an i.i.d.~fashion from a sufficiently large finite field $\mathbb{F}_q^L$,
\begin{align}
H(W_k)&=L, \quad   k \in \{1, \cdots, K\}\\
H(\cw)&=H(W_1, W_2, \cdots, W_K)=KL,\quad (q\text{-ary symbols})
\end{align}

\begin{figure}[t]
	\centering
	\includegraphics[width=0.4\textwidth]{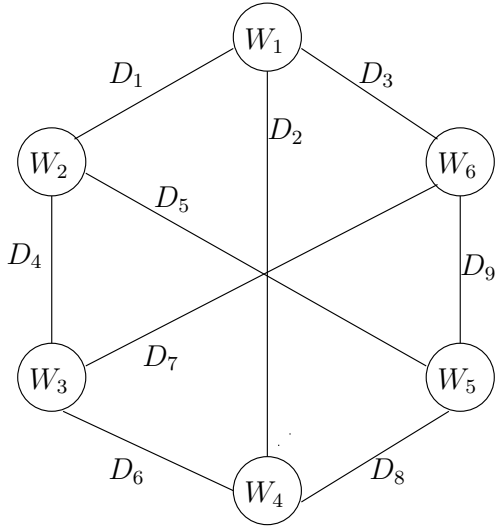}
	\caption{Graph structure for an example $(6,3,2,9)$ storage system.}
	\label{graph_example}
\end{figure}

The storage system is parameterized by $(K,R,M,N)$. We note that for a feasible storage system (that is symmetric across databases and messages), we have $KR=MN$. In this work, we focus on the case $M=2$. To fully characterize the storage system in this case, we represent the storage system as \emph{$R$-regular graph}\footnote{We note that the graph used in our formulation may be considered as the dual graph of the one used in \cite{TamoISIT}. In our work, $M=2$, the nodes are the messages and the edges are the databases, while in \cite{TamoISIT}, $R=2$, the nodes are the databases and the edges are the messages.}; see Fig.~\ref{graph_example} and Table~\ref{storage(6,3,2,9)} for a $(6,3,2,9)$ example. We characterize the storage system by a $(V,E)$ regular graph, where $V=\cw=\{W_1, W_2, \cdots, W_K\}$ is the set of vertices, and $E=\mathcal{D}=\{D_1, D_2, \cdots, D_N\}$ is the set of edges, i.e., in this graph, the vertices are the messages and the edges are the databases. An edge $D_j$ drawn between messages $W_m$ and $W_k$ means that the contents of database $D_j$ is $Z_j=\{W_m, W_k\}$. This graph is an $R$-regular graph, since each message is repeated $R$ times across the storage system. In the following, we define specific parameters of the graph, which are needed while constructing the converse proof.

\begin{definition}[Graph reduction] \label{defn-reduce}
	The graph $(V,E)=(\cw,\mathcal{D})$ is reduced iteratively starting with the vertex $W_1$ by enumerating all the edges connecting to $W_1$, and removing all neighboring vertices connected to enumerated edges except one, which we denote by $\tilde{W}_2$. The process of enumerating edges and removing corresponding neighbors iteratively continues until one vertex is left $\tilde{W}_{\kappa+1}$ after $\kappa$ reductions.
\end{definition}

\begin{definition}[Spread of the graph]\label{spread}
	The spread of a graph $\delta$ is the largest sequence of edges (databases) that results from the graph reduction procedure given in Definition~\ref{defn-reduce}.
\end{definition}

\begin{table}[t]
	\centering
	\begin{tabular}{|c|c|c|c|c|c|c|c|c|}
		\hline
		$D_1$& $D_2$ & $D_3$ & $D_4$ & $D_5$ & $D_6$ & $D_7$ & $D_8$ & $D_9$ \\
		\hline \hline
		$W_1$& $W_1$  & $W_1$ & $W_2$ & $W_2$ & $W_3$ & $W_3$ & $W_4$ & $W_5$  \\
		\hline
		$W_2$& $W_4$  & $W_6$ & $W_3$ & $W_5$ & $W_4$ & $W_6$ & $W_5$ & $W_6$ \\
		\hline
	\end{tabular}
\caption{Contents of databases for the example $(6,3,2,9)$ system specified by graph in Fig.~\ref{graph_example}.}
\label{storage(6,3,2,9)}
\end{table}

\begin{figure}[t]
	\centering
	\begin{subfigure}[b]{0.48\textwidth}
		\centering
		\includegraphics[width=0.7\textwidth]{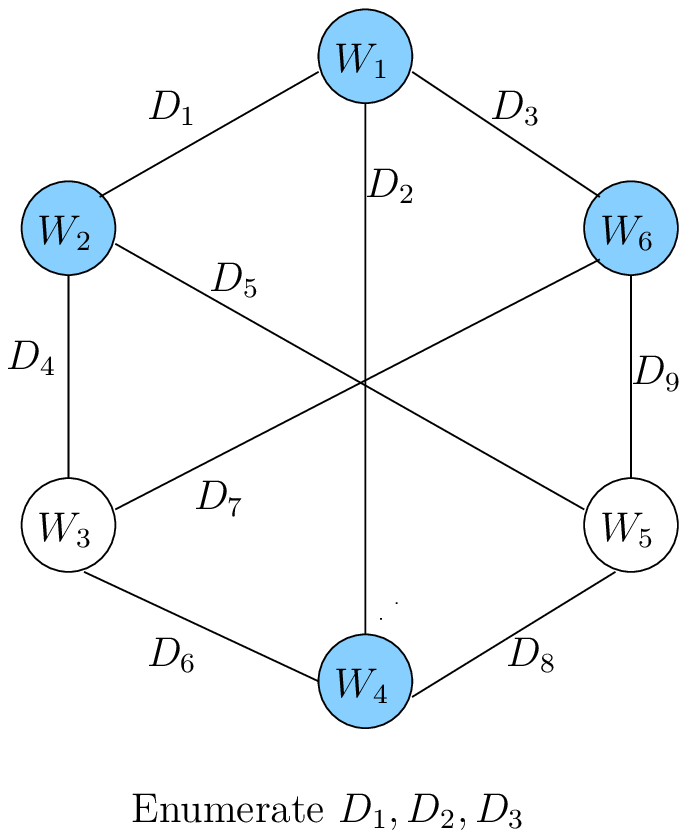}
		\caption{}
	\end{subfigure}
	\begin{subfigure}[b]{0.48\textwidth}
		\centering
		\includegraphics[width=0.7\textwidth]{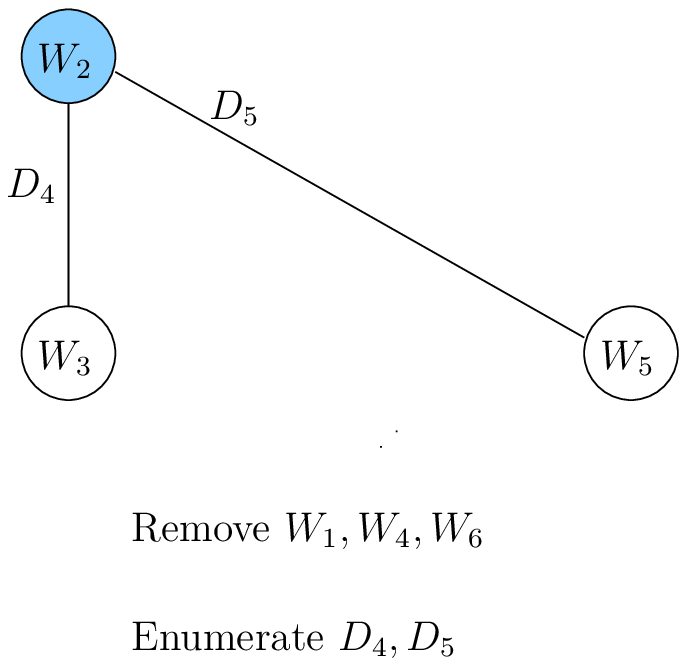}
		\caption{}
	\end{subfigure}
	\caption{Graph reduction for the example $(6,3,2,9)$ storage system. The end result of graph reduction is the sequence of databases (edges) $D_1,D_2,D_3,D_4,D_5$. (a) Neighbors of $W_1$ are specified and the connected databases are enumerated. (b) All neighboring vertices except $W_2$ are removed, and so on.}
	\label{graph_reduction}
\end{figure}

An example graph reduction for the $(6,3,2,9)$ storage system given in Fig.~\ref{graph_example} and Table~\ref{storage(6,3,2,9)} is shown in Fig.~\ref{graph_reduction}. In this work, we further focus on two special classes of $R$-regular graphs, namely: cyclic graphs and fully-connected graphs.

\begin{definition}[Cyclic graphs]
	The graph $(V,E)=(\cw,\mathcal{D})$ is called a cyclic graph if each two adjacent vertices are connected by an edge and no non-adjacent vertices are connected by an edge, i.e., the contents of the databases can be written as (without loss of generality):
	\begin{align}
	Z_1&=\{W_1,W_2\} \nonumber \\
	Z_2&=\{W_2,W_3\} \nonumber \\
	\vdots \nonumber\\
	Z_{N-1}&=\{W_{N-1},W_N\}\nonumber \\
	Z_N&=\{W_N,W_1\}
	\end{align}
	Consequently, the cyclic graph is parameterized by $(K,R,M,N)=(K,2,2,K)$, and the spread of the graph is $\delta=K-1$.
\end{definition}

\begin{definition}[Fully-connected graphs]
	The graph $(V,E)=(\cw,\mathcal{D})$ is called fully-connected if every two vertices are connected by a unique edge. Hence, the contents of the databases can be written as the $\binom{K}{2}$ subsets of $\{1, \cdots, K\}$ with 2 elements. The fully-connected graph is parameterized by $(K,R,M,N)=(K,K-1,2,\binom{K}{2})$, and the spread of the graph is $\delta=K-1$.
\end{definition}

In PIR, the user wants to retrieve a message $W_k$ without leaking any information about the identity of the message to any individual database. To that end, the user sends $N$ queries, one for each database. These queries are independent of the messages as the user has no information about the messages prior to retrieval, hence,
\begin{align}\label{independence}
I(\cw;Q_{1:N}^{[k]})=0, \quad k \in [K]
\end{align}

The databases respond to the user queries by answer strings $A_{1:N}^{[k]}$. The answer string $A_n^{[k]}$ is a deterministic function of the query $Q_n^{[k]}$ and the contents of the database $D_n$, which is denoted by $Z_n$, therefore,
\begin{align}\label{answer_string}
H(A_n^{[k]}|Q_n^{[k]}, Z_n)=0, \quad n \in [N]
\end{align}

In PIR, we have two formal requirements. First, we have the privacy requirement. To ensure privacy, the retrieval strategy intended to retrieve $W_i$ must be indistinguishable from the retrieval strategy intended to retrieve $W_j$ for any $i$ and $j$, i.e.,
\begin{align}\label{privacy}
(Q_n^{[i]},A_n^{[i]},\cw) \sim (Q_n^{[j]},A_n^{[j]},\cw), \quad n \in [N], \: i,j \in [K]
\end{align}
where $\sim$ denotes statistical equivalence.

The second requirement is the reliability requirement. The user needs to be able to reconstruct $W_k$ perfectly\footnote{The results of this work do not change if we relaxed the reliability constraint to allow arbitrarily small probability of error, i.e., if we changed the reliability constraint as $H(W_k|Q_{1:N}^{[k]}, A_{1:N}^{[k]})=o(L)$.} from the collected answers, i.e.,
\begin{align}\label{reliability}
H(W_k|Q_{1:N}^{[k]}, A_{1:N}^{[k]})=0
\end{align}

We measure the efficiency of the retrieval scheme by the retrieval rate $R_{\text{PIR}}$. An achievable retrieval scheme is a scheme that satisfies \eqref{privacy}, \eqref{reliability} for some message length $L$. The retrieval rate is the ratio between the length of the desired message $L$ and the total download,
\begin{align}
R_{\text{PIR}}=\frac{L}{\sum_{n=1}^{N} H(A_n^{[k]})}
\end{align}
The PIR capacity is the largest PIR rate over all achievable schemes, i.e., $C_{\text{PIR}}=\sup R_{\text{PIR}}$.

\section{Main Results}
In this section, we present the main results of this paper. Our first result is a general upper bound for storage systems defined by $R$-regular graphs with $M=2$ and arbitrary $(K,R,N)$ which is given in the following theorem. The proof of Theorem~\ref{Thm_upper} is given in Section~\ref{sect:converse}.

\begin{theorem}[Upper-bound for $R$-regular graphs]\label{Thm_upper}
	For an $R$-regular graph storage system with $(K,R,M,N)=(K,R,2,N)$, the retrieval rate is upper bounded by
	\begin{align}
	R_{\text{PIR}} \leq \min \left\{\frac{R}{N}, \frac{1}{1+\frac{\delta}{R}}\right\}
	\end{align}
\end{theorem}

\begin{remark}
	The upper bound reveals a dependency on the structure of the storage system, captured in the spread of the graph $\delta$. I.e., the upper bound cannot be parameterized by $(K,R,M,N)$ only. This opens the door for joint optimization of the storage system together with the retrieval scheme.
\end{remark}

\begin{remark}\label{remark4}
	The upper bound $R_{\text{PIR}} \leq \frac{1}{1+\frac{\delta}{R}}$ is a general upper bound which is valid for any storage system with $M=2$ and is represented via an $R$-regular graph (including the example shown in Figs.~\ref{graph_example} and \ref{graph_reduction}). In this paper, we focus on two special cases, namely:
	\begin{itemize}
	\item Cyclic graphs: In this case, the spread of the graph is $\delta=K-1$ as we can cover all the messages in the storage system by visiting exactly $K-1$ databases. Furthermore, $R=2$, as every node in the graph is connected to $2$ adjacent nodes only. Applying the bound in Theorem~\ref{Thm_upper}, $R_{\text{PIR}} \leq \frac{1}{1+\frac{\delta}{R}}= \frac{1}{1+\frac{K-1}{2}}=\frac{2}{K+1}$.
	\item Fully-connected graphs: In this case, the spread of the graph is $\delta=R$ as $W_1$ is connected to all other messages. Applying the bound in Theorem~\ref{Thm_upper}, $R_{\text{PIR}} \leq \frac{1}{1+\frac{\delta}{R}}=\frac{1}{1+\frac{R}{R}}=\frac{1}{2}$. Also, in this case, $R=K-1$ and $N=K(K-1)/2$, hence, $R_{\text{PIR}} \leq \frac{R}{N}=\frac{(K-1)}{K(K-1)/2}=\frac{2}{K}$.
	\end{itemize}
\end{remark}

In the following two results, we characterize the PIR capacity of cyclic graphs and fully-connected graphs. The converse proofs for Theorems~\ref{Thm2} and \ref{Thm3} are corollaries of Theorem~\ref{Thm_upper} as shown in Remark~\ref{remark4}. The achievability proofs of Theorems~\ref{Thm2} and \ref{Thm3} are given in Section~\ref{achievability}.

\begin{theorem}[Capacity of cyclic graphs]\label{Thm2}
	For a cyclic graph storage system, the PIR capacity is given by
	\begin{align}
	C_{\text{PIR}}=\frac{2}{K+1}
	\end{align}
\end{theorem}

\begin{theorem}[Capacity of fully-connected graphs]\label{Thm3}
	For a fully-connected graph storage system with $M=2$, the PIR capacity is given by
	\begin{align}
	C_{\text{PIR}}=\left\{
	\begin{array}{ll}
	\frac{1}{2}, \quad & K=2,3 \\
	\frac{2}{K}, \quad & K \geq 4 \\
	\end{array}
	\right.
	\end{align}
\end{theorem}

\begin{remark}
	The capacity results in this work reveal a severe loss in the retrieval rate due to non-replication. For the cyclic and fully-connected graphs, $C_{\text{PIR}} \rightarrow 0$ as $N \rightarrow \infty$. This is in contrast to the classical PIR problem \cite{JafarPIR}, where $C_{\text{PIR}} \rightarrow 1$ as $N \rightarrow \infty$. This is intuitively due to the fact that as $N \rightarrow \infty$, we have $K \rightarrow \infty$. Meanwhile, the number of side information equations generated is limited due to non-replication. In particular, the side information equations are related to $R-1$ (in contrast to $N-1$ in the classical model), while total downloads grow with $N$ as the user needs to download from all databases to satisfy the privacy constraint. The ratio $\frac{R}{N} \rightarrow 0$ as $N \rightarrow \infty$ for both cases.
\end{remark}

\begin{remark}
	The results of this work outperform the trivial scheme of downloading all messages, which achieves $\frac{1}{K}$. Our retrieval rate also outperforms the best achievable scheme in \cite{TamoISIT}, which achieves $\frac{1}{N}$ for cyclic graphs (which is the comparable case to our work). The achievable rate $\frac{1}{N}=\frac{1}{K}<\frac{2}{K+1}$ for the case of cyclic graphs. This implies that the retrieval rates in non-replicated PIR systems in \cite{TamoISIT} may be improved. Nevertheless, the results in \cite{TamoISIT} are more general which are valid for all $(K,2,M,N)$ graph-based storage systems. The results in \cite{TamoISIT} also cover collusion resistance, which is outside the scope of our work here.
\end{remark}


\section{Converse Proof} \label{sect:converse}
In this section, we prove Theorem~\ref{Thm_upper}. To that end, we present a general upper bound for the retrieval rate for general $R$-regular graphs for the case of $M=2$.

Let $\cq$ denote the collection of all queries to all databases for all desired messages, i.e.,
\begin{align}
\cq \triangleq \left\{Q_n^{[k]}: k \in [K],\: n \in [N]\right\}
\end{align}
We assume that the retrieval scheme is symmetric across databases (as in \cite[Lemma~1]{PIR_Globecomm}). This assumption is without loss of generality, since any asymmetric retrieval scheme can be transformed into a symmetric one by means of time-sharing without changing the retrieval rate. Hence, for $m \in \{1, \cdots, K\}$, we have
\begin{align}\label{symmetry}
H(A_1^{[m]}|\cq)&=H(A_n^{[m]}|\cq), \quad n \in \{1, \cdots, N\}\\
H(A_1^{[m]}|\cw\setminus\{W_m\},\cq)&=H(A_n^{[m]}|\cw\setminus\{W_m\},\cq), \quad n \in \{1, \cdots, N\}\label{symmetry1}
\end{align}
where $\cw\setminus\{W_m\}=\{W_1, W_2, \cdots, W_{m-1},W_{m+1}, \cdots, W_K\}$. We need the following lemma.

\begin{lemma}\label{lemma_ILB}
	Let $\cR_m$ denote the set of databases containing message $W_m$, then
	\begin{align}
	H(A_n^{[m]}|\cw\setminus\{W_m\},\cq) \geq \frac{L}{R}, \quad n \in \cR_m
	\end{align}
\end{lemma}

\begin{Proof}
We have
\begin{align}
L&=H(W_m)\\
 &=H(W_m|\cw\setminus\{W_m\},\cq)  \label{lemma_ILB1}\\
 &=H(W_m|\cw\setminus\{W_m\},\cq)-H(W_m|\cw\setminus\{W_m\},\cq, A_{1:N}^{[m]})  \label{lemma_ILB2}\\
 &=H(W_m|\cw\setminus\{W_m\},\cq)-H(W_m|\cw\setminus\{W_m\},\cq, A_{\cR_m}^{[m]}) \label{lemma_ILB3}\\
 &=I(W_m;A_{\cR_m}^{[m]}|\cw\setminus\{W_m\},\cq)\\
 &=H(A_{\cR_m}^{[m]}|\cw\setminus\{W_m\},\cq) \label{lemma_ILB4}\\
 &\leq RH(A_n^{[m]}|\cw\setminus\{W_m\},\cq), \quad n \in \cR_m \label{lemma_ILB5}
\end{align}
where \eqref{lemma_ILB1} follows from the fact that $W_m$ is independent of the messages and the queries $(\cw\setminus\{W_m\},\cq)$, \eqref{lemma_ILB2} follows from the reliability constraint. For \eqref{lemma_ILB3}, we note that the answer strings $A_{[N]\setminus \cR_m}^{[m]}=\{A_n^{[m]}: n \not\in \cR_m \}$ are deterministic functions of $(\cw\setminus\{W_m\},\cq)$ only, hence $W_m \rightarrow (\cw\setminus\{W_m\},\cq) \rightarrow A_{[N]\setminus \cR_m}^{[m]}$ is a Markov chain and $A_{[N]\setminus \cR_m}^{[m]}$ can be dropped from the conditioning. \eqref{lemma_ILB4} follows from the fact that answer strings are deterministic functions of the messages and queries, and \eqref{lemma_ILB5} follows from the database symmetry in \eqref{symmetry1}. Rearranging \eqref{lemma_ILB5} concludes the proof.
\end{Proof}

We are now ready to prove the converse statement in Theorem~\ref{Thm_upper}. We first prove that $R_{\text{PIR}} \leq \frac{R}{N}$. From Lemma~\ref{lemma_ILB}, we have
\begin{align}
L&\leq RH(A_n^{[m]}|\cw\setminus\{W_m\},\cq), \quad n \in \cR_m \\
 &=\frac{R}{N} NH(A_n^{[m]}|\cw\setminus\{W_m\},\cq)\\
 &\leq \frac{R}{N} NH(A_{n}^{[m]}|\cq) \label{conv1_1}\\
&=\frac{R}{N} \sum_{n=1}^{N}H(A_n^{[m]}|\cq) \label{conv1_2}
\end{align}
where \eqref{conv1_1} follows from the fact that conditioning reduces entropy, \eqref{conv1_2} follows from the symmetry across databases. Therefore,
\begin{align} \label{thm1-proof1}
R_{\text{PIR}} = \frac{L}{\sum_{n=1}^{N} H(A_n^{[m]})} \leq \frac{L}{\sum_{n=1}^{N}H(A_n^{[m]}|\cq)} \leq \frac{R}{N}
\end{align}


Next, we prove that $R_{\text{PIR}} \leq \frac{1}{1+\frac{\delta}{R}}$, where $\delta$ is the spread of the graph; see Definition~\ref{spread}. In order to obtain the spread of the graph, we begin by the node representing $W_1$, then we enumerate all the edges (databases) connecting to $W_1$. Without loss of generality, label these databases by $1, 2 , \cdots, R$. These edges are connecting to the nodes corresponding to $\{W_{n_1}, W_{n_2}, \cdots, W_{n_R}\}$. Then, we reduce the graph by removing all the connecting nodes to $W_1$ except one (which belongs to the path of the largest distance), which we denote by $\tilde{W}_2 \in \{W_{n_1}, W_{n_2}, \cdots, W_{n_R}\}$. We again enumerate all the edges connecting to $\tilde{W}_2$ with the nodes $\{W_{n_{R+1}}, W_{n_{R+2}}, \cdots, W_{n_{R+\delta_2}}\}$, where $\delta_2$ is the number of databases that contain $\tilde{W}_2$ after the graph reduction, then we reduce the graph again by removing all nodes connecting to $\tilde{W}_2$ except one, which we denote by $\tilde{W}_3$, and so on. Then, we have
\begin{align}
L&=H(W_1) \\
 &=H(W_1|\cq)-H(W_1|A_{1:N}^{[1]},\cq) \label{ub1}\\
 &=I(W_1;A_{1:N}^{[1]}|\cq) \\
 &=H(A_{1:N}^{[1]}|\cq)-H(A_{1:N}^{[1]}|W_1,\cq) \\
 &\leq NH(A_1^{[1]}|\cq)-H(A_{1:N}^{[1]}|W_1,\cq) \label{ub2}\\
 &\leq NH(A_1^{[1]}|\cq)-H(A_{\Delta}^{[1]}|W_1,\cq) \label{ub3}\\
 &=NH(A_1^{[1]}|\cq)-\sum_{i=1}^{\delta} H(A_{i}^{[1]}|W_1,\cq,A_{1:i-1}^{[1]})\\
 &\leq NH(A_1^{[1]}|\cq)-\sum_{i=1}^{\delta} H(A_{i}^{[1]}|W_1, \cw\setminus\{W_{n_i}\},\cq,A_{1:i-1}^{[1]}) \label{ub4}\\
 &=NH(A_1^{[1]}|\cq)-\sum_{i=1}^{\delta} H(A_{i}^{[1]}|W_1, \cw\setminus\{W_{n_i}\},\cq) \label{ub5}
\end{align}
where \eqref{ub1} follows from the reliability constraint and the independence of queries and messages, \eqref{ub2} follows from the independence bound, \eqref{ub3} follows from the non-negativity of the entropy function where $A_{\Delta}^{[1]}$ denotes the answer strings returned by the sequence of the databases that define the spread of the graph, and \eqref{ub4} follows from the fact that conditioning on $\cw\setminus\{W_i\}$ cannot increase entropy.
	
To show \eqref{ub5}, we note that from the reduction procedure that results in  $A_{\Delta}^{[1]}$, we have
\begin{align}
A_i^{[1]}=\left\{
\begin{array}{ll}
f_i(\cq,W_1,W_{n_i}), \quad & 1\leq i \leq R \\
f_i(\cq,\tilde{W}_2,W_{n_i}), \quad &R+1 \leq i \leq R+\delta_2 \\
f_i(\cq,\tilde{W}_3,W_{n_i}), \quad &R+\delta_2+1 \leq i \leq R+\delta_2+\delta_3 \\
\vdots\\
f_i(\cq,\tilde{W}_\kappa,W_{n_i}), \quad &R+\sum_{j=2}^{\kappa-1} \delta_j+1 \leq i \leq R+\sum_{j=2}^{\kappa} \delta_j	
\end{array}
\right.
\end{align}
for some deterministic function $f_i(\cdot)$, and $\kappa$ is the number of reductions on the graph until all nodes are removed from the graph. Since the leading message at the $j$th graph reduction $\tilde{W}_j$ belongs to the set of the connected messages in the $(j-1)$th graph reduction, and at the $j$th graph reduction, the nodes connecting to $\tilde{W}_j$ are removed from the graph, we have $\{\tilde{W}_2, \tilde{W}_3, \cdots, \tilde{W}_{j(i)}\}\subseteq\{W_{n_1}, W_{n_2}, \cdots, W_{n_{i-1}}\} \subseteq \cw \setminus \{W_{n_i}\}$, where $j(i)$ is the index of the leading message in the $i$th database. Consequently, we can drop $A_{1:i-1}^{[1]}$ as they are deterministic functions of $(\cq,W_1,\cw \setminus \{W_{n_i}\})$.
	
Now, we have
\begin{align}
L&\leq NH(A_1^{[1]}|\cq)-\sum_{i=1}^{\delta} H(A_{i}^{[1]}|W_1, \cw\setminus\{W_{n_i}\},\cq)\\
 &=NH(A_1^{[1]}|\cq)-\sum_{i=1}^{\delta} H(A_{i}^{[n_i]}|W_1, \cw\setminus\{W_{n_i}\},\cq) \label{ub6}\\
 &\leq NH(A_1^{[1]}|\cq)-\sum_{i=1}^{\delta} \frac{L}{R} \label{ub7}\\
 &=NH(A_1^{[1]}|\cq)- \frac{\delta L}{R}
\end{align}
where \eqref{ub6} follows from the privacy constraint, and \eqref{ub7} follows from Lemma~\ref{lemma_ILB}. Reordering terms, we have
\begin{align}
R_{\text{PIR}} = \frac{L}{\sum_{n=1}^{N} H(A_n^{[m]})} \leq \frac{L}{NH(A_1^{[1]}|\cq)} \leq \frac{1}{1+\frac{\delta}{R}}
\end{align}
which together with (\ref{thm1-proof1}) concludes the proof of Theorem~\ref{Thm_upper}.

\section{Achievability Proof}\label{achievability}
In this section, we begin first with a motivating example of $(K,R,M,N)=(3,2,2,3)$ to show the basic ingredients of the achievable scheme. In fact, the graph for this motivating example is both cyclic and fully-connected (see Fig.~\ref{motivating_example_graph}), therefore, this motivating example can be considered as a unifying instance of the optimal scheme for both cyclic and fully-connected graphs. Then, we present general capacity-achieving schemes for cyclic graphs and fully-connected graphs. Finally, we show by an example how we can extend the presented schemes to the case of $M \geq 3$ (with no claim of optimality).

\subsection{Motivating Example: $K=3$, $R=2$, $M=2$, $N=3$}\label{motivating}
\begin{figure}[t]
	\centering
	\includegraphics[width=0.35\textwidth]{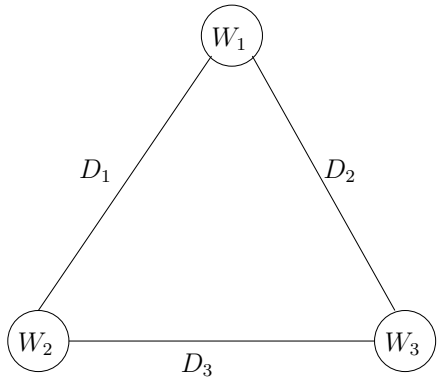}
	\caption{Graph structure for the $(3,2,2,3)$ system used as a motivating example.}
	\label{motivating_example_graph}
\end{figure}

In this example, we consider a storage system that consists of $N=3$ databases. The system stores $K=3$ messages in total, namely $W_1, W_2, W_3$. Each message is replicated across $R=2$ databases, such that each database stores $M=2$ messages (see Table~\ref{motivating_example}). This is a cyclic and also a fully-connected graph as shown in Fig.~\ref{motivating_example_graph}.

\begin{table}[t]
	\centering
	\begin{tabular}{|c|c|c|}
		\hline
		Database~1 ($D_1$)& Database~2 ($D_2$) & Database~3 ($D_3$) \\
		\hline \hline
		$W_1$& $W_1$  & $W_2$ \\
		\hline
		$W_2$& $W_3$  & $W_3$ \\
		\hline
	\end{tabular}
\caption{Contents of databases for the $(3,2,2,3)$ system specified by graph in Fig.~\ref{motivating_example_graph}.}
\label{motivating_example}
\end{table}


Without loss of generality, assume that the desired message is $W_1$. To construct the capacity-achieving scheme, the user randomly permutes the indices of messages $W_1, W_2, W_3$ independently, uniformly, and privately from the databases. Denote the permuted version of $W_1$ by the vector $(a_1,\cdots, a_L)$, the permuted version of $W_2$ by $(b_1, \cdots, b_L)$, and the permuted version of $W_3$ by $(c_1, \cdots, c_L)$. Pick $L=12$.

A straightforward solution for this problem is to apply Sun and Jafar scheme in \cite{JafarPIR}. Since every database contains $M=2$ messages, the user downloads a single bit from each message from each database in round 1, i.e., the user downloads $a_1, b_1$ from database 1, $a_2, c_1$ from database 2, and $b_2,c_2$ from database 3. Now, the user exploits $b_2, c_2$ as side information by downloading $a_3+b_2$ from database 1, and $a_4+c_2$ from database 2. Finally, the user downloads the sum $b_3+c_3$ from database 3. The query table for this scheme is shown in Table~\ref{table_Sun_Jafar}. Note that although the sum $b_3+c_3$ is irrelevant to the decodability of $W_1$, the user needs to download it to satisfy the privacy constraint. Otherwise, database 3 would figure out that the desired message is $W_1$, as the user requests 2 bits from database 3 when the desired message is $W_1$, while the user would have requested 3 bits from database 3 if the desired message was $W_2$ or $W_3$. With this scheme, the user downloads $4$ bits from $W_1$ out of the total $9$ downloads, hence $R_{\text{PIR}}=\frac{4}{9}$.

\begin{table}[h]
	\centering
	\begin{tabular}{|c|c|c|}
		\hline
		Database 1 & Database 2 & Database 3 \\
		\hline
		$a_1$ & $a_2$ & $b_2$ \\
		$b_1$ & $c_1$ & $c_2$ \\
		\hline
		$a_3+b_2$ & $a_4+c_2$ & $b_3+c_3$ \\
		\hline
	\end{tabular}
\caption{Sun and Jafar scheme for the $K=3$, $R=2$, $M=2$, $N=3$ example.}
\label{table_Sun_Jafar}
\end{table}

Although this scheme outperforms the scheme in \cite{TamoISIT} in terms of the retrieval rate (the scheme in \cite{TamoISIT} achieves $R_{\text{PIR}}=\frac{1}{3}$), there is room for improving it. The main source of inefficiency of the scheme is the downloads from database~3, as the user downloads 3 bits and exploits only 2 of them. Moreover, the user downloads new independent bit $b_3+c_3$. If the user introduces \emph{dependency} to the downloads of database~3, the user may \emph{compress}\footnote{Throughout this work, we use the expressions ``dependency'' and ``compression''. In previous PIR works, the user downloads new and independent undesired symbols at each round, which can be used in later rounds as side information. However, in this work, the user downloads undesired symbols which are \emph{dependent} on the undesired symbols downloaded from other databases. We download these dependent symbols even from the databases that do not contain the desired message. We call these ``dependent'' downloads to differentiate them from ``side information'' downloads, which are intended to be used to decode the desired message directly. Furthermore, by ``compression'', we mean downloading shorter (fewer) answer strings than the greedy algorithm in \cite{JafarPIR} by exploiting the knowledge of the dependent symbols.} the requests from database~3, and improve the retrieval rate. In order to do this, the user downloads the sums $b_1+c_2$ and $b_2+c_1$ from database~3 (see Table~\ref{Sun_Jafar_compress}). For the decodability, the user can decode $c_2$ by canceling $b_1$ from $b_1+c_2$ and $b_2$ by canceling $c_1$ from $b_2+c_1$. Therefore, $a_3$, $a_4$ are decodable by canceling $b_2$ and $c_2$.

\begin{table}[h]
	\centering
	\begin{tabular}{|c|c|c|}
		\hline
		Database 1 & Database 2 & Database 3 \\
		\hline
		$a_1$ & $a_2$ &  \\
		$b_1$ & $c_1$ &  \\
		\hline
		$a_3+b_2$ & $a_4+c_2$ & $b_1+c_2$ \\
		& & $b_2+c_1$\\
		\hline
	\end{tabular}
\caption{Compressing the scheme of Sun and Jafar for $K=3$, $R=2$, $M=2$, $N=3$.}
\label{Sun_Jafar_compress}
\end{table}

Nevertheless, the scheme in Table~\ref{Sun_Jafar_compress} is \emph{not private} because the user still downloads 2 bits from database~3 in the form of sum of 2 bits. To remedy this problem, the user should repeat the compression of the downloads over all databases, i.e., the user should download 2 bits in the same manner of downloading from database~3 in the other two databases as well. Hence, in repetition~2, the user compresses the downloads from database~2 and downloads $a_7+c_3$, $a_8+c_4$. Similarly, in repetition~3, the user downloads $a_{10}+b_5$ and $a_{11}+b_6$ from database 1. The complete query structure is given in Table~\ref{complete_motivating_example}.

\begin{table}[t]
	\centering
	\begin{tabular}{|c|c|c|c|}
		\hline
		& Database 1 & Database 2 & Database 3 \\
		\hline
		\multirow{4}{*}{{\rotatebox[origin=c]{90}{\parbox[c]{2.1cm}{\centering rep.~1}}}} &$a_1$ & $a_2$ &  \\
		&$b_1$ & $c_1$ &  \\
		\cline{2-4}
		&$a_3+b_2$ & $a_4+c_2$ & $b_1+c_2$ \\
		& & & $b_2+c_1$\\
		\hline
		\hline
		\multirow{4}{*}{{\rotatebox[origin=c]{90}{\parbox[c]{2.1cm}{\centering rep.~2}}}} &$a_5$ &  & $b_4$ \\
		&$b_3$ &  & $c_3$ \\
		\cline{2-4}
		&$a_6+b_4$ & $a_7+c_3$ & $b_3+c_4$ \\
				 & & $a_8+c_4$ & \\
		\hline
		\hline
		\multirow{4}{*}{{\rotatebox[origin=c]{90}{\parbox[c]{2.1cm}{\centering rep.~3}}}} &  & $a_9$ & $b_5$ \\
		& & $c_5$ & $c_6$ \\
		\cline{2-4}
		&$a_{10}+b_5$ &$a_{12}+c_6$  & $b_6+c_5$  \\
		&$a_{11}+b_6$ &  & \\
		\hline
	\end{tabular}
\caption{Complete query structure for the capacity-achieving scheme for $K=3$, $R=2$, $M=2$, $N=3$.}
\label{complete_motivating_example}
\end{table}

Next, we discuss privacy, decodability and the rate of this achievable scheme.

Regarding privacy: The query structure is now symmetric across the databases, and the indices of the bits from each message are chosen uniformly, independently and privately. Hence, all queries are equally likely, and the scheme is private.

Regarding decodability: We note that each repetition is decodable separately. As we discussed above, $a_1,\cdots,a_4$ are decodable in repetition~1. For repetition~2, $a_5$ is decodable directly, $a_6$ is decodable by canceling $b_4$ from $a_6+b_4$, and $a_7$ is decodable by canceling $c_3$ from $a_7+c_3$. Finally, $c_4$ is decodable by canceling $b_3$ from $b_3+c_4$ and therefore $a_8$ is decodable by further canceling $c_4$ from $a_8+c_4$ (or equivalently by adding $a_8+c_4$ and $b_3+c_4$ under modulo-2 addition and canceling $b_3$ from the sum). The decodability of repetition~3 follows in a similar way to the decodability of repetition~2 by exchanging the roles of $W_2, W_3$.

Regarding the achievable rate: The user downloads $12$ bits from $W_1$ out of a total of $24$ downloads. Consequently, $R_{\text{PIR}}=\frac{12}{24}=\frac{1}{2}$ which matches the upper bound in Theorem~\ref{Thm_upper}.

\begin{remark}
	It is interesting to compare the PIR capacity here to the PIR capacity in \cite{StorageConstrainedPIR} where the contents are stored in the databases using the optimal storage strategy under the memory-size constraint $\mu$. Note that, in this example, $\mu=\frac{2}{3}$ as every database stores 2 full messages out of 3 messages. Using the optimal storage strategy in \cite{StorageConstrainedPIR}, the PIR capacity is $C_{\text{PIR}}=(1+\frac{1}{2}+\frac{1}{2^2})^{-1}=\frac{4}{7}$ which is larger than the PIR capacity here $C_{\text{PIR}}=\frac{1}{2}$. This implies a loss in the PIR capacity due to storing full messages here as opposed to storing uncoded parts of the messages in \cite{StorageConstrainedPIR} subject to the same memory-size constraint.
\end{remark}

\subsection{General Achievability for the Case of Cyclic Graphs}
In this section, we generalize the ideas of the motivating example for arbitrary $K$. The new ingredient in this scheme (in contrast to \cite{JafarPIR}) is the \emph{compression} of the queries submitted for a subset of the databases. To satisfy the privacy constraint, the user performs the scheme along $\binom{N}{2}=\binom{K}{2}$ repetitions. In each repetition, the user chooses to submit the full query (according to \cite{JafarPIR}) to $2$ databases. For the remaining databases, the user downloads two symbols in the form of 2-sums. The scheme works with $L=4\binom{K}{2}$ symbols. The general scheme for cyclic graphs can be summarized as:

\begin{enumerate}
	\item \emph{Index preparation:} The indices of the symbols of each message are permuted independently, uniformly, and privately at the user side.
	\item \emph{Constructing full queries:} We apply the scheme of Sun and Jafar \cite{JafarPIR} to construct the full queries to all databases. We apply this scheme over blocks of $\tilde{L}=4$. To that end, the user downloads 1 individual symbol from each message from each database in round 1. Next, the user downloads a 2-sum from the stored messages in each database. This sum exploits the side information generated from other databases. Note that since $R=2$ in this graph, the user can generate 1 side information equation for each database. Another change from \cite{JafarPIR} is that even for the $K-2$ databases that do not contain the desired message, the user exploits the side information generated at other databases by introducing dependency to the answers.
	\item \emph{Compressing queries:} The user choose different $K-2$ databases at each repetition. The user compresses the queries to these databases by adding the individual symbols in round 1 into single equation.\footnote{We note that in some cases, we may need to shuffle the indices of the symbols in the sum to prevent ending up with useless equations. For example, if the full queries are in the form of $b_2,c_1,b_1+c_2$, then, after compressing, we have the 2-sums $b_2+c_1$, and $b_1+c_2$. Now, imagine that $b_2$ and $c_1$ are decodable from the remaining databases. In this case, the sum of $b_2+c_1$ is useless and the sum $b_1+c_2$ is not decodable. However, if we shuffle the indices such that the user downloads $b_2+c_2$ and $b_1+c_2$, then the user can use both equations to decode $b_1$ and $c_2$. This would not affect the privacy as the indices are permuted uniformly and privately at the user side.}
	\item Repeat step 2, 3 over new blocks of 4 symbols for $\binom{K}{2}$ repetitions.
\end{enumerate}

\subsubsection{Decodability, Privacy, and Achievable Rate}
Regarding decodability: In this scheme, at each repetition, we have $4$ unknowns corresponding to the desired message and $2(K-1)$ unknowns corresponding to the undesired messages. The user downloads 3 equations (full queries) from 2 databases, and 2 equations from the remaining databases. Hence, the user downloads in total $6+2(K-2)=2K+2$ equations in $2(K-1)+4$ unknowns. This linear system is decodable (up to necessary index shuffling).

Regarding privacy: The scheme is private since the symbols are permuted randomly and privately at the user side and the scheme is repeated along all $\binom{K}{2}$ combinations of the databases. Hence, the structure of the queries is the same across all databases. Thus, the distribution of the queries is the same irrespective to the desired message.

Regarding the achievable rate: From every repetition of the scheme, the user can decode 4 symbols from the desired message, thus,
\begin{align}
R_{\text{PIR}}=\frac{4}{2(K-2)+3*2}=\frac{2}{K+1}
\end{align}

\subsection{General Achievability for the Case of Fully-Connected Graphs}
In this section, we present the general achievability for the case of fully-connected graphs. For $K=2$, we have 1 database containing 2 messages; the capacity-achieving scheme is simply to download the contents of the entire database, hence $C_{\text{PIR}}=\frac{1}{2}$. For $K=3$, the capacity-achieving scheme is exactly the motivating example in Section~\ref{motivating}, hence $R_{\text{PIR}}=\frac{1}{2}$.

For $K \geq 4$, the upper bound $R_{\text{PIR}} \leq \frac{2}{K}$ is the active upper bound. The general achievability for this case is given below. The achievable scheme works with $L=R=K-1$ symbols from $\mathbb{F}_q$, where $q$ is sufficiently large and is prime.

\begin{enumerate}
	\item \emph{Index preparation:} The indices of the symbols of each message is permuted independently, uniformly, and privately at the user side.
	\item \emph{Retrieval from database~1:} Denote the permuted contents of the $n$th database by $Z_n=\left\{X_1^{(n)},X_2^{(n)}\right\}$. Without loss of generality, assume that the desired message is stored in database~1, hence, $X_1^{(1)}$ is the permuted version of the desired message. From database~1, the user downloads a weighted sum of two symbols from the two messages, i.e., the user downloads $\alpha_1^{(1)} X_1^{(1)}(1)+\alpha_2^{(1)} X_2^{(1)}(1)$ from database~1, where $\alpha_m^{(n)}  \in \mathbb{F}_q$, $m \in \{1,2\}, n \in [N]$. The choice of $\alpha_m^{(n)}$ will be specified later.
	\item \emph{Exploiting side information:} The user downloads different weighted sums from every database. If the $n$th database contains the desired message, the user downloads a new desired symbol in the sum. If the message stored in the $n$th database is undesired, the user exploits the same message symbol in all databases. I.e., the user downloads the weighted sum $\alpha_1^{(n)} X_1^{(n)}(i)+\alpha_2^{(1)} X_2^{(1)}(j)$, where indices $i,j$ are chosen depending on the message (if desired, we increment the index; if undesired we fix the index to 1)
	\item \emph{Database symmetry:} The user repeats the last step across all databases.
\end{enumerate}

\subsubsection{Decodability, Privacy, and Achievable Rate}
Regarding decodability: The user collects $N=\binom{K}{2}$ equations. These equations have $L=K-1$ unknowns corresponding to the desired message and $K-1$ unknowns corresponding to the undesired messages, i.e., we have a linear system of $\binom{K}{2}$ equations in $2K-2$ unknowns. Without loss of generality, assume that the desired message is stored in the first $K-1$ databases, hence $W_k=\{X_1^{(1)}(1),X_1^{(2)}(2), \cdots, X_1^{(K-1)}(K-1)\}$. The linear system of equations can be written as:
\begin{align}
\underbrace{\begin{bmatrix}
\alpha_1^{(1)} & 0 & \cdots & 0 & \alpha_2^{(1)} & 0 & 0 & \cdots & 0 \\
 0 & \alpha_1^{(2)} & \cdots & 0 & 0 & \alpha_2^{(2)}& 0 & \cdots & 0 \\
 \vdots &  \vdots &  \vdots &  \vdots &  \vdots &  \vdots&  \vdots&  \vdots &  \vdots \\
 0 & 0 & \cdots & \alpha_1^{(K-1)} & 0 & 0 & 0 & \cdots & \alpha_2^{(K-1)} \\
 0 & 0 & \cdots & 0 & \alpha_1^{(K)} & \alpha_2^{(K)} & 0 & \cdots & 0 \\
 \vdots &  \vdots &  \vdots &  \vdots &  \vdots &  \vdots&  \vdots&  \vdots &  \vdots \\
  0 & 0 & \cdots & 0 & 0 & 0  & \cdots & \alpha_1^{\binom{K}{2}} & \alpha_2^{\binom{K}{2}} \\
\end{bmatrix}}_{\mathbf{\Psi}}
\begin{bmatrix}
X_1^{(1)}(1) \\
X_1^{(2)}(2) \\
\vdots \\
X_1^{(K-1)} (K-1) \\
X_2^{(1)}(1) \\
X_2^{(2)}(1) \\
\vdots \\
X_2^{(K-1)}(1)
\end{bmatrix}
=\begin{bmatrix}
A_1^{[k]} \\
A_2^{[k]} \\
\vdots\\
A_{K-1}^{[k]}\\
A_{K}^{[k]}  \\
A_{K+1}^{[k]}  \\
\vdots \\
A_{\binom{K}{2}}^{[k]}  \\
\end{bmatrix}
\end{align}
The choice of the coefficients $\alpha_m^{(n)}  \in \mathbb{F}_q$, $m \in \{1,2\}, n \in [N]$ is such that the decoding matrix $\mathbf{\Psi}$ is invertible. One simple way\footnote{In general, one can enumerate all possible $\mathbf{\Psi}$ that are full rank for every desired message $W_k$. The user can choose $\mathbf{\Psi}$ uniformly from this set.} to choose $\boldsymbol{\alpha}=\left(\alpha_1^{(1)}, \alpha_2^{(1)}, \alpha_1^{(2)}, \cdots, \alpha_{2}^{\binom{K}{2}}\right)$ is to choose $\boldsymbol$ uniformly from all possible $P(q,2 \binom{K}{2})=\frac{q!}{(q-2 \binom{K}{2})!}$ permutations of the field elements.

Regarding privacy: Since the message symbols and the coefficients are permuted uniformly, and the distribution of the queries for every database is the same irrespective of the desired message, the retrieval scheme is private.

Regarding the achievable rate: The user downloads $N=\binom{K}{2}$ answer strings, $L=K-1$ of which are desired symbols and are decodable, thus,
\begin{align}
R_{\text{PIR}}=\frac{K-1}{\binom{K}{2}}=\frac{2}{K}
\end{align}

\subsubsection{Further Example: Fully-Connected Graph with $K=4$}
\begin{figure}[t]
	\centering
	\includegraphics[width=0.4\textwidth]{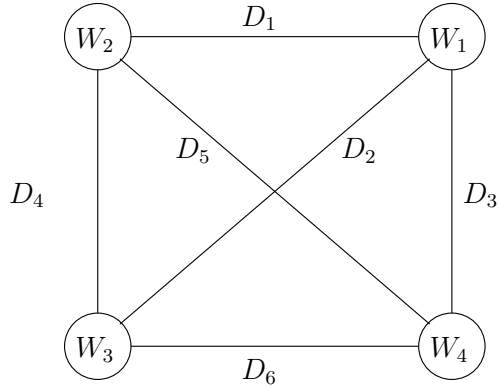}
	\caption{Graph structure for the $(4,3,2,6)$ system which is fully-connected with $K=4$.}
	\label{FC_K4_figure}
\end{figure}

As a concrete example, we present the achievable scheme for $K=4$ for a fully-connected graph. Hence, we have $N=\binom{K}{2}=6$ databases and $R=K-1=3$. Thus, this is a $(4,3,2,6)$ system. We assume that $Z_1=\{W_1,W_2\}$, $Z_2=\{W_1,W_3\}$, $Z_3=\{W_1,W_4\}$, $Z_4=\{W_2,W_3\}$, $Z_5=\{W_2,W_4\}$, and $Z_6=\{W_3,W_4\}$. See Fig.~\ref{FC_K4_figure} and Table~\ref{table_FC_4} for the graph structure and the database contents. Assume for sake of simplicity that $\boldsymbol{\alpha}=(1,2,\cdots,12) \in \mathbb{F}_{13}^{12}$. The scheme works with $L=3$ bits. Denote the permuted message $W_1$ by $(a_1,a_2,a_3)$, $W_2$ by $(b_1,b_2,b_3)$, and so on. Assume without loss of generality that the user is interested in retrieving $W_1$. Therefore, the user downloads the following:
\begin{align}
A_1^{[1]}&=a_1+2b_1 \\
A_2^{[1]}&=3a_2+4c_1 \\
A_3^{[1]}&=5a_3+6d_1 \\
A_4^{[1]}&=7b_1+8c_1 \\
A_5^{[1]}&=9b_1+10d_1\\
A_6^{[1]}&=11c_1+12d_1
\end{align}
This system of equations is full-rank, hence $W_1$ is decodable along with $b_1,c_1,d_1$. The rate of retrieval is $R_{\text{PIR}}=\frac{3}{6}=\frac{1}{2}$. To see privacy, let the desired message be $W_2$, in which case, the user downloads:
\begin{align}
A_1^{[2]}&=a_1+2b_1 \\
A_2^{[2]}&=3a_1+4c_1 \\
A_3^{[2]}&=5a_1+6d_1 \\
A_4^{[2]}&=7b_2+8c_1 \\
A_5^{[2]}&=9b_3+10d_1\\
A_6^{[2]}&=11c_1+12d_1
\end{align}
This system of equations is full rank as well. Since, the queries have the same structure and the symbol indices are chosen randomly and privately, the scheme is private.

\begin{table}[t]
	\centering
	\begin{tabular}{|c|c|c|c|c|c|}
		\hline
		$D_1$& $D_2$ & $D_3$ & $D_4$ & $D_5$ & $D_6$ \\
		\hline  \hline
		$W_1$& $W_1$  & $W_1$ & $W_2$ & $W_2$ & $W_3$\\
		\hline
		$W_2$& $W_3$  & $W_4$ & $W_3$ & $W_4$ & $W_4$\\
		\hline
	\end{tabular}
\caption{Contents of databases for the $(4,3,2,6)$ system specified by graph in Fig.~\ref{FC_K4_figure}}
\label{table_FC_4}
\end{table}

\subsection{Discussion and Further Extensions: Extension to $M \geq 3$: }
In this section, we show how the ideas of $M=2$ can be extended to $M \geq 3$. We discuss our additional ideas via the following example. In this example, we consider a $(K,R,M,N)=(6,2,3,4)$ storage system, whose structure is shown in Table~\ref{table_M3}. Note that, in this example, our graph formulation fails to represent the storage structure since $M>2$, however, we can use the graph structure in \cite{TamoISIT} as $R=2$ in this example. In the following, we only show an achievable scheme for this example without any claim of optimality. In addition to introducing dependency as in the previous schemes, we have a new insight in this case, which is to exploit \emph{processed side information}.

\begin{table}[h]
	\centering
	\begin{tabular}{|c|c|c|c|}
		\hline
		Database~1 ($D_1$)& Database~2 ($D_2$) & Database~3 ($D_3$) & Database~4 ($D_4$)  \\
		\hline  \hline
		$W_1$& $W_1$  & $W_2$ & $W_3$\\
		\hline
		$W_2$& $W_4$  & $W_3$ & $W_4$\\
		\hline
		$W_5$& $W_6$ & $W_6$ & $W_5$ \\
		\hline
	\end{tabular}
	\caption{Contents of databases for the $(6,2,3,4)$ system. Note that, here $M=3$.}
	\label{table_M3}
\end{table}

Our scheme works with $L=18$ symbols from $\mathbb{F}_2$. We permute the indices of the messages uniformly, independently and privately at the user side. We denote the permuted message symbols of $W_1, W_2, \cdots, W_6$ by the vectors $(a_1, \cdots, a_{L}), (b_1, \cdots, b_L), \cdots, (f_1, \cdots, f_L)$.

The idea of our scheme is to extend the greedy algorithm of \cite{JafarPIR} to our setting (see Table~\ref{achievable_M3}). To that end, the user starts by downloading $2$ individual symbols from every message from every database in round~1. Hence, the user downloads $a_1,a_2$, $b_1,b_2$, $e_1,e_2$ from database~1, $a_3,a_4$, $d_1,d_2$, $f_1,f_2$ from database~2, and so on.

Returning to the scheme in \cite{JafarPIR}, the user downloads 2-sums in round~2 from all databases. The undesired symbols in round~1 are exploited as side information in round~2. In our case, this is applicable in databases~1 and 2. Therefore, the user downloads the sums $a_5+b_3$, $a_6+b_4$, $a_7+e_3$, $a_8+e_4$ from database~1, and $a_9+d_3$, $a_{10}+d_4$, $a_{11}+f_3$, $a_{12}+f_4$. We complete round~2 by downloading $b_5+e_5$, $b_6+e_6$ from database~1, and $d_5+f_5$, $d_6+f_6$ from database~2 to satisfy the privacy constraint by downloading all combinations of the 2-sums.

In order to proceed with the achievable scheme, we need to generate side information in round~3, which consists of 3-sums. At this point, we note two issues: First, we did not exploit the side information generated from $W_3$ in round~1, as the messages $W_1,W_3$ do not appear together at any database. Second, we note that the side information needed in round~3 does not appear directly in any other database unlike \cite{JafarPIR}, i.e., in round~3, we need the side information to be of the form $b+e$ and $d+f$, which are not available in any other database. This motivates the use of \emph{processed side information}, i.e., combine side information generated at multiple databases into a single side information that is usable at another database. There are two types of processing in this example, which are: combining double 2-sums and combining triple 2-sums.

First, for combining double 2-sums to get a single side information equation, we download $b_7+c_7$ from database~3 and $c_7+e_7$ from database~4. By adding the two 2-sums (modulo-2 addition), we get $b_7+e_7$ which can be used as side information in database~1. Similarly, we obtain the single side information $b_8+e_8$ by adding $b_8+c_8$ and $c_8+e_8$ and again use it in database~1. Next, we generate the side information needed in database~2. We combine $c_9+f_7$ from database~3 with $c_9+d_7$ from database~4 to get $d_7+f_7$, and combine $c_5+f_8$ from database~3 and $c_5+d_8$ from database~4 to get $d_8+f_8$.

Second, we can create extra side information by combining triple 2-sums. To see that, we can add $b_5+e_5$ from database~1, $b_5+f_9$ from database~3, and $d_9+e_5$ from database~4 to create the side information $d_9+f_9$, which can be exploited in database~2. Similarly, we can add $d_5+f_5$ from database~2, $b_9+f_5$ from database~3, and $d_5+e_9$ from database~4 to get $b_9+e_9$, which can be exploited in database~1.

To introduce dependency in databases~3 and 4 as in the previous schemes, we can download $b_1+c_3+f_1$, $b_2+c_4+f_2$, $c_1+d_1+e_1$, and $c_2+d_2+e_2$, which result from round~1.

Using this scheme, the user gets $18$ desired symbols out of total $60$ downloads, resulting in $R_{\text{PIR}}=\frac{18}{60}=\frac{3}{10}$. This outperforms the achievable scheme of \cite{TamoISIT}, which achieves $\frac{1}{N}=\frac{1}{4}$.

We note that this is the first instance of using processed side information in PIR. Further, the presented scheme achieves the bound in Lemma~\ref{lemma_ILB} with equality, i.e., $H(A_n^{[m]}|\cw\setminus\{W_m\},\cq) = \frac{L}{R}=9$, which may be promising. However, a curious question remains which should be investigated further, which is: Can we compress the downloads in the same manner of the achievable scheme for the cyclic graphs by exploiting dependencies?

\begin{table}[t]
	\centering
	\begin{tabular}{|c|c|c|c|}
		\hline
		Database 1 & Database 2 & Database 3 & Database 4\\
		\hline
		$a_1,a_2$ & $a_3,a_4$ & $b_3,b_4$ & $c_3,c_4$\\
		$b_1,b_2$ & $d_1,d_2$ & $c_1,c_2$ & $d_3,d_4$\\
		$e_1,e_2$ & $f_1,f_2$ & $f_3,f_4$ & $e_3,e_4$\\
		\hline
		$a_5+b_3$ & $a_9+d_3$ & $b_7+c_7$ & $c_9+d_7$\\
		$a_6+b_4$ & $a_{10}+d_4$ & $b_8+c_8$ & $c_5+d_8$\\
		$a_7+e_3$ & $a_{11}+f_3$ & $b_9+f_5$ & $c_7+e_7$\\
		$a_8+e_4$ & $a_{12}+f_4$ & $b_5+f_9$ & $c_8+e_8$\\
		$b_5+e_5$ & $d_5+f_5$ & $c_9+f_7$ & $d_5+e_9$ \\
		$b_6+e_6$ & $d_6+f_6$ & $c_5+f_8$ & $d_9+e_5$ \\
		\hline
		$a_{13}+b_7+e_7$ & $a_{16}+d_7+f_7$& $b_1+c_3+f_1$ & $c_1+d_1+e_1$ \\
		$a_{14}+b_8+e_8$ & $a_{17}+d_8+f_8$& $b_2+c_4+f_2$ & $c_2+d_2+e_2$ \\
		$a_{15}+b_9+e_9$ & $a_{18}+d_9+f_9$& $b_6+c_6+f_6$ & $c_6+d_6+e_6$ \\
		\hline
	\end{tabular}
	\caption{An achievable scheme for the $(6,2,3,4)$ storage system.}
	\label{achievable_M3}
\end{table}

\section{Conclusion}
In this paper, we investigated the PIR problem from non-replicated and non-colluding databases. We studied the $(K,R,2,N)$ storage systems, where every database stores $M=2$ messages. This system is uniquely described by an $R$-regular graph. We proved a general upper bound, which depends on the spread of the graph. We derived the capacity of two classes of graphs, namely: cyclic graphs and fully-connected graphs. For these two classes of graphs, we proposed novel achievable schemes, whose retrieval rate matches the developed upper bound. Our results showed that non-replication significantly hurts the retrieval rate.

\bibliographystyle{unsrt}
\bibliography{references_new}
\end{document}